\documentclass[epjCONF,columns]{svjour}
\usepackage{graphics}
\usepackage[varg]{txfonts} 
\usepackage[latin1]{inputenc}
\usepackage{lineno}
\session-title{Hadron Collider Physics Symposium 2011}
\begin{document}
%
\title{Search for flavor-changing neutral currents in top quark decays at ATLAS}
\author{ A. Cortes Gonzalez on behalf of the ATLAS Collaboration\thanks{\email{arely.cortes.gonzalez@cern.ch}} }
\institute{University of Illinois at Urbana-Champaign}
%
\abstract{
ATLAS results are presented on a search for flavor-changing neutral currents (FCNC) in top quark decays at the LHC.  Events are searched for where a pair of top quarks is produced and one decays through the Standard Model $t \rightarrow Wb$ mode while the other decays through the FCNC mode  $t \rightarrow qZ$.  Leptonic final states are used in which both the $W$- and $Z$-bosons decay leptonically, yielding final states with three leptons and two jets.  The observed events are consistent with the expected backgrounds from Drell-Yan and diboson production and a limit on the FCNC branching ratio of top quarks to $qZ$ is set.
} 
\maketitle
%

\section{Introduction}
\label{intro}
If anomalous top-quark couplings exist, this would affect production and/or decay processes at hadron colliders, such as the LHC. A rare top-quark decay is that which occurs via FCNC. These decays are highly suppressed in the Standard Model (SM), consequently observation of such a decay would be suggestive of new physics. \\

A search was done for $t\bar{t}$ pairs with one top quark decaying through flavor-changing neutral currents ($t\rightarrow qZ$) and the other through the SM dominant mode ($t \rightarrow bW$). Only the leptonic decays of the $W$- and the $Z$-bosons are considered signal. The main background sources are $ZZ$ and $WZ$ events, which include three charged leptons in the final state, and were estimated with Monte Carlo (MC) simulation. Backgrounds with one, two, or three fake leptons, were estimated by data-driven (DD) methods \cite{bib:main}.

\section{Data samples}
\label{datasamples}
The search for FCNC top quark signals was done in $\sqrt{s} = 7$ TeV $pp$-collision data taken by the ATLAS experiment between March and June 2011. Only the periods in which all the sub-detectors were operational were considered, resulting in a data sample corresponding to a total integrated luminosity of 0.70 fb$^{-1}$, with an estimated uncertainty of 3.7\% \cite{bib:lumi}.

\section{Monte Carlo simulation samples}
\label{MC}
MC simulation samples of top quark pair production,
with one of the top quarks decaying through FCNC to a Z boson and the other top quark with a SM decay, were generated with TopReX \cite{bib:toprex}.

\section{Object definition}
\label{objects}
After offline reconstruction, extra requirements were applied to the reconstructed objects. \\

Muons were required to have $p_T > 20 $~GeV, $|\eta| < $ 2.5 and a distance $\Delta R > 0.4$ to the closest reconstructed jet with $p_T > 20$~GeV. Only muons with a good track quality were considered. Isolation requirements were also applied to the candidate muons: the sum of the transverse momenta of the tracks inside a cone of size $\Delta R = 0.3$ around the muon had to be less than 4 GeV and the energy deposited inside a cone of size $\Delta R = 0.3$ around the muon (ignoring its energy) to be less than 4 GeV.\\ 

Electrons were required to be isolated such that the energy deposited inside a cone of size $\Delta R = 0.2$ around their direction was less than 3.5 GeV. Electrons were additionally required to have $E_T > 20 $~GeV. Only electrons with $|\eta| < 2.47$, excluding the crack region $1.37 < |\eta| < 1.52$, were selected, and jets with their axis within $\Delta R < 0.2$ from the electron direction were removed.\\

Jets were reconstructed with the anti-kt algorithm ($R = 0.4$) starting from topological energy clusters in the calorimeter at the electromagnetic scale appropriate for the energy deposited by electrons or photons \cite{bib:jets}. Jets were then calibrated with MC simulation-based $p_T$- and $\eta$-dependent correction factors to restore the full hadronic energy scale. Only jets with $p_T > 20 $~GeV were considered.\\

The missing transverse energy ($E_T^{miss}$) was calculated \cite{bib:met} from topological clusters calibrated at the electromagnetic scale and corrected according to the energy scale of the associated object. Muons, which are not primarily measured by the calorimeter, were included in the $E_T^{miss}$ calculation using their momentum measured in the tracking and muon spectrometer systems.\\

\section{Selection, reconstruction, and analysis of events}
\label{selection}

\subsection{Event preselection}
\label{preselection}
\begin{figure}
\centering
\resizebox{0.9\columnwidth}{!}{%
   \includegraphics{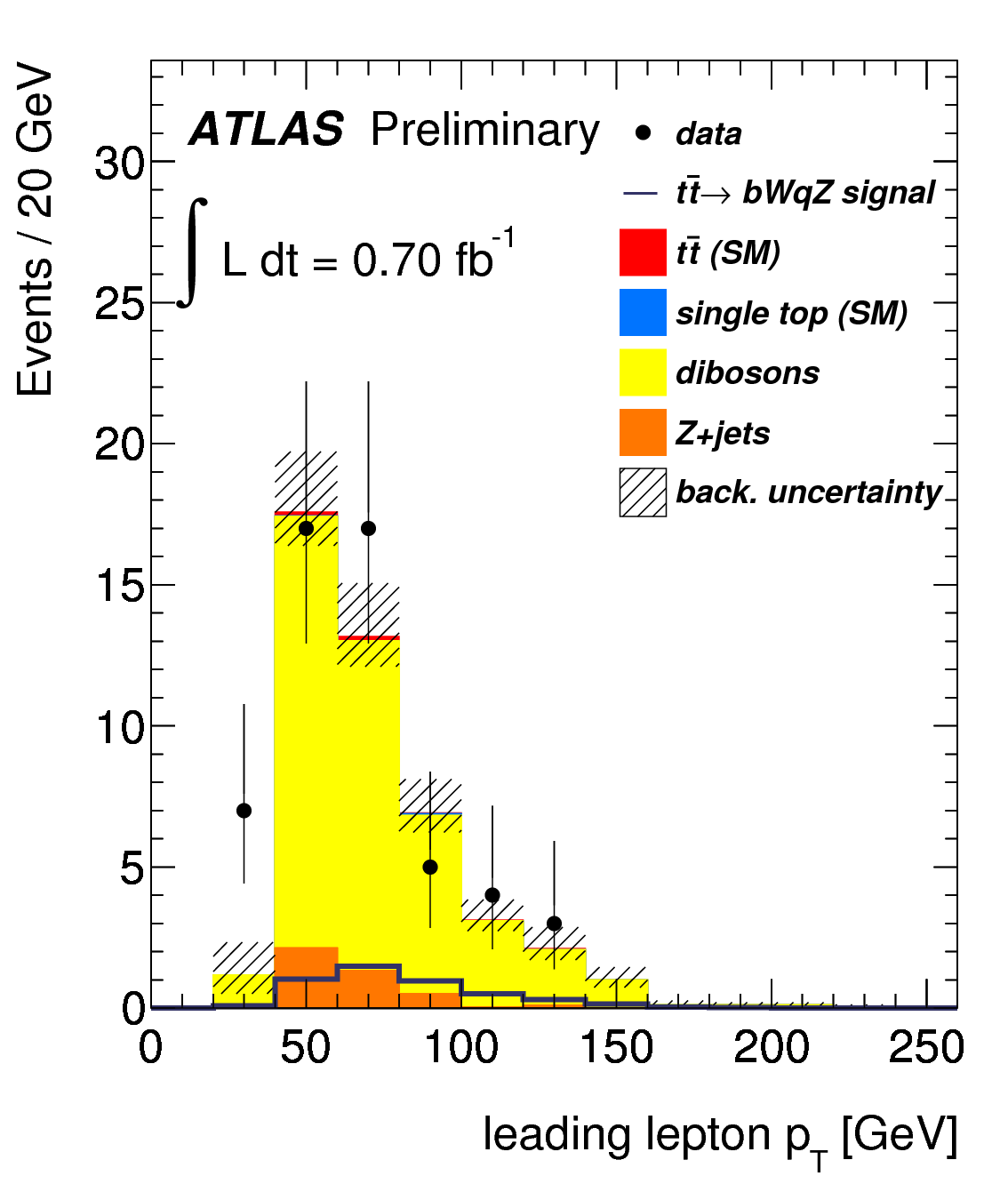} 
   \includegraphics{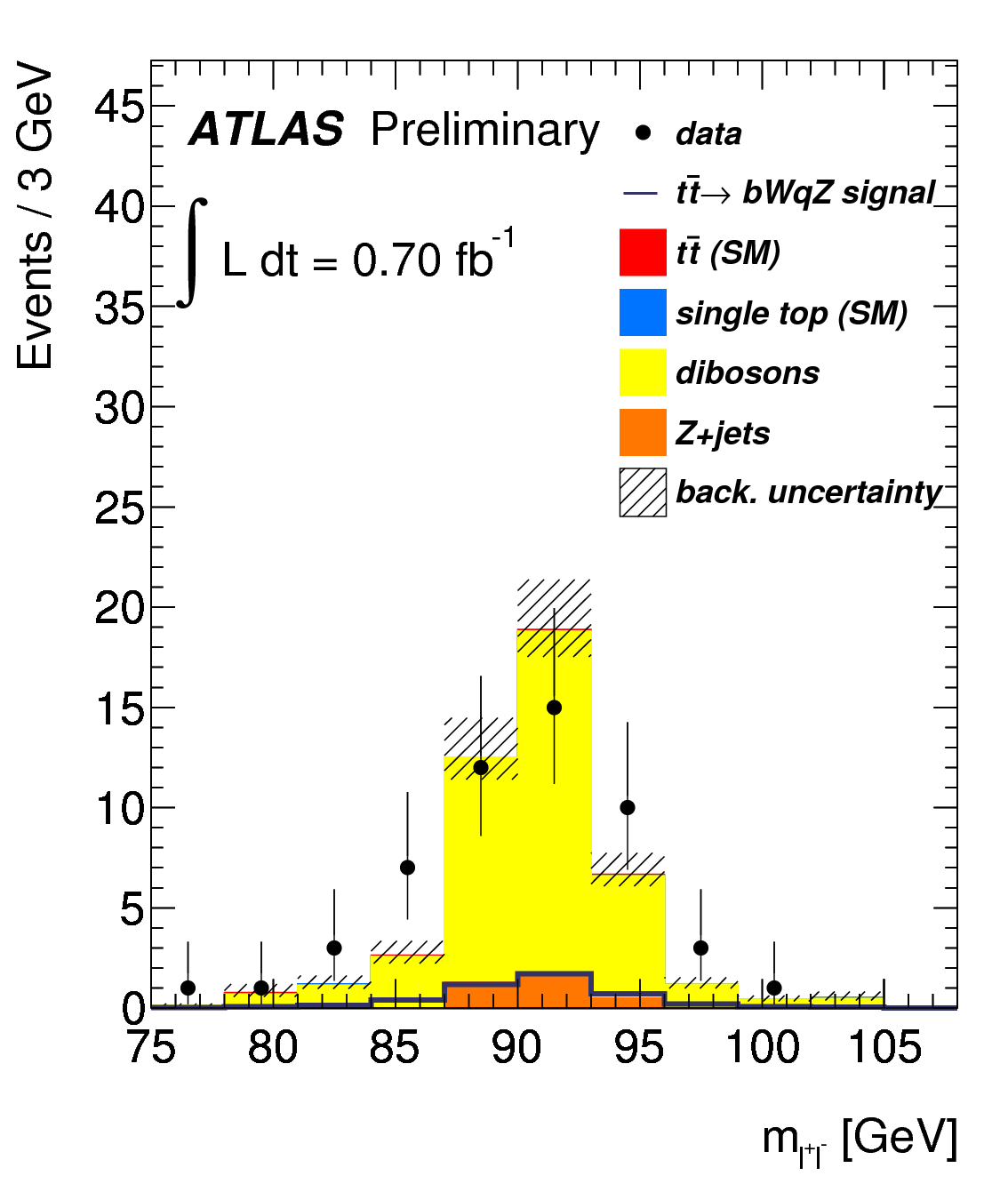}   
   }
\resizebox{0.9\columnwidth}{!}{%
   \includegraphics{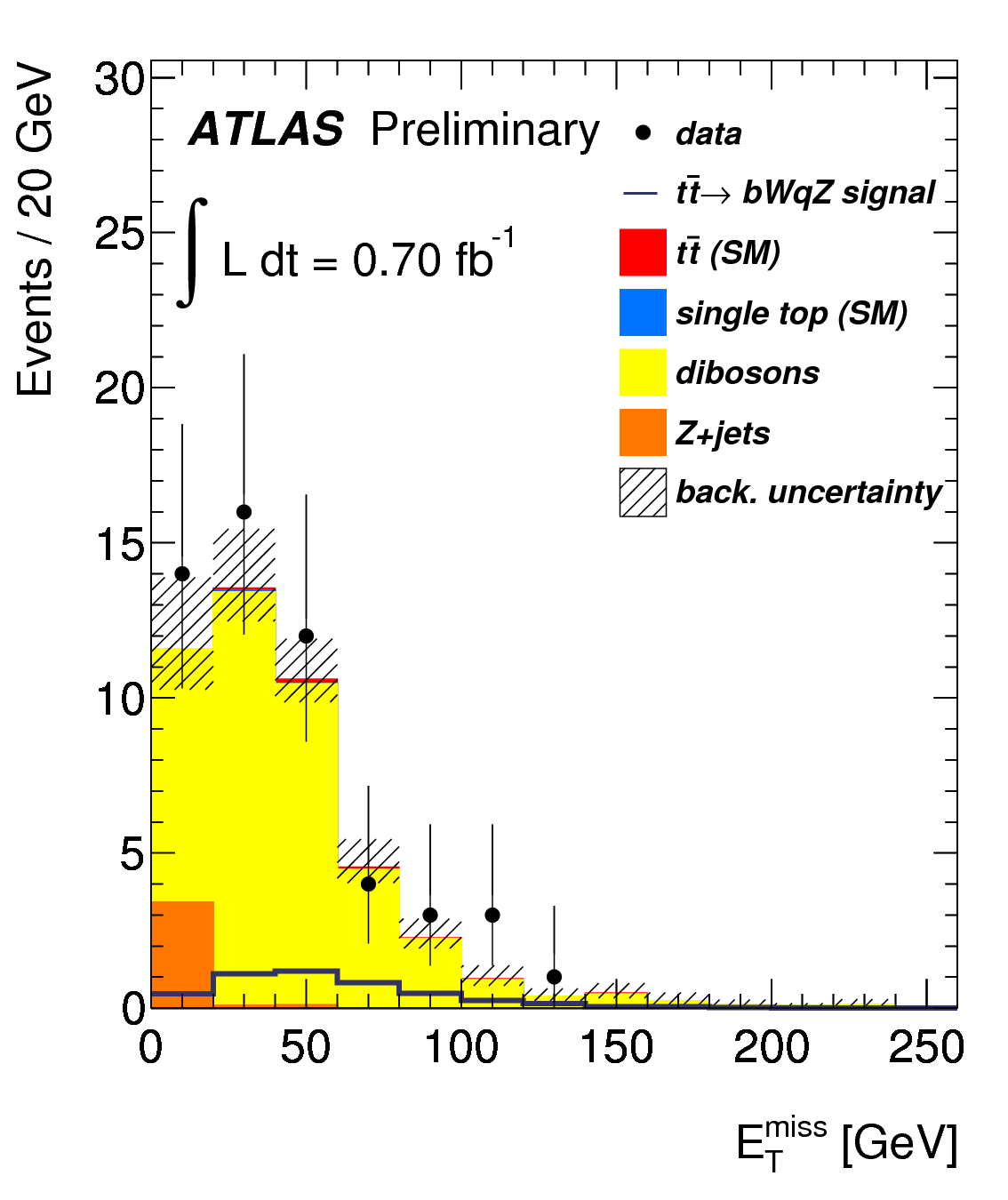}
   \includegraphics{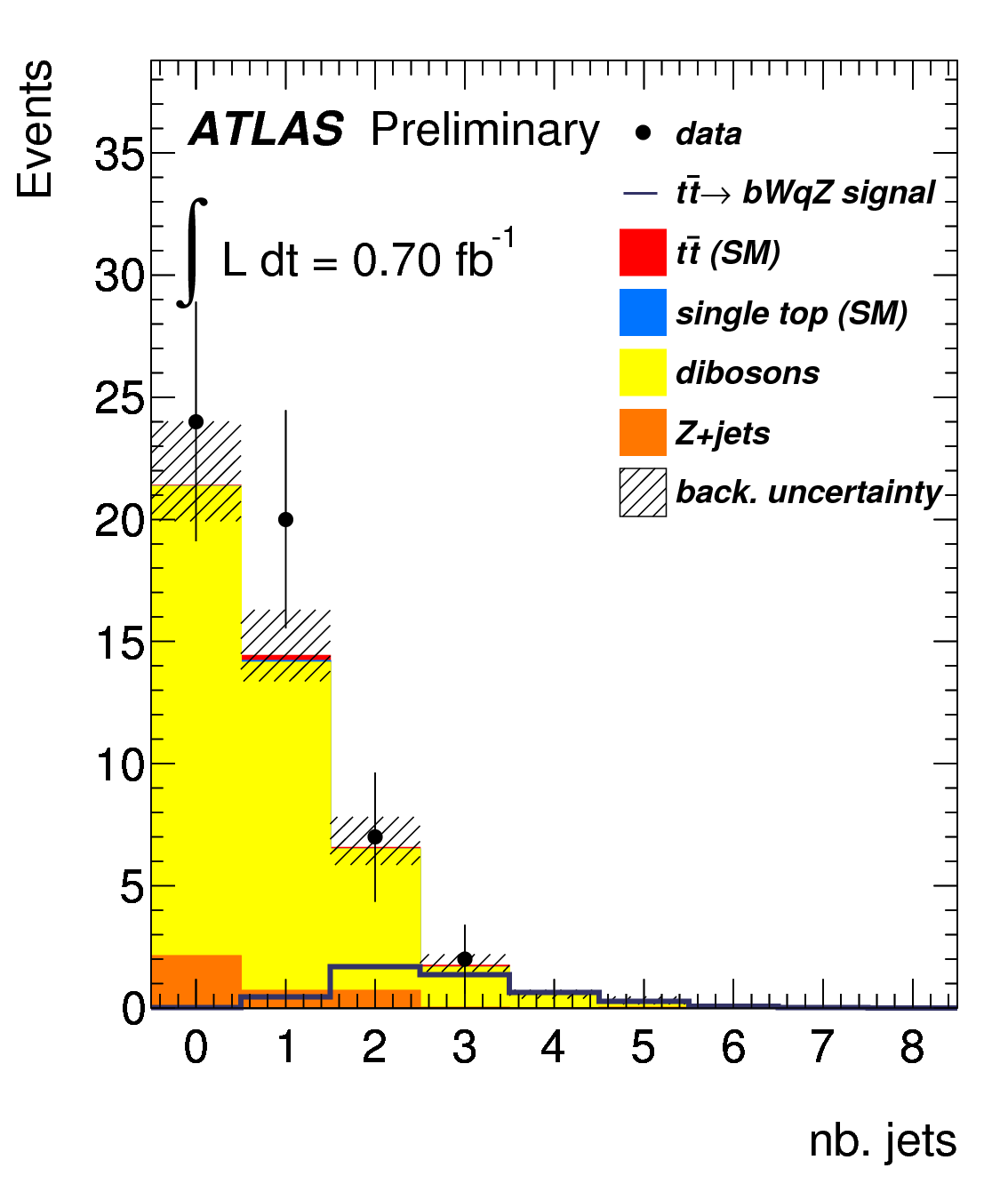}  
}
\caption{Distributions obtained with the $t \rightarrow qZ$ analysis after the preselection. Transverse momentum of the leading lepton (top left), reconstructed mass of the two leptons with same flavor and opposite charge (top right); $E_T^{miss}$ (bottom left) and number of jets (bottom right), are shown. The background uncertainties shown include the MC simulation statistical uncertainties and the DD uncertainties. The signal distributions are normalized to the observed BR limit, at 95\% CL.}
\label{fig:preselection}  
\end{figure}

Events were required to pass the single lepton trigger (electron with $p_T > 20$~GeV or muon with $p_T>18$~GeV), and have at least one primary vertex with more than four tracks. Events with three isolated leptons were selected, with all three leptons matched to the same primary vertex. The transverse momentum of the leading lepton and the other two leptons were required to be $>25$~GeV and $>20$~GeV, respectively, with $|\eta| < 2.47$ ($|\eta|<2.5$) for electrons (muons). Furthermore, events were required to contain at least two leptons of the same flavor and opposite charge. The preselection was completed by requiring that the reconstructed mass of one lepton pair with the same flavor and opposite charges should be within 15~GeV of the known $Z$-mass peak. Distributions of the relevant variables obtained after these cuts are shown in Figure~\ref{fig:preselection}. \\
%

\subsection{Event reconstruction and final selection}
\label{reconstruction}
The event selection was completed by requiring at least two jets, with $|\eta| < 2.5$; the transverse momentum of the leading (sub-leading) jet had to be $>30$~GeV ($>20$~GeV) and $E_T^{miss}>20$~GeV. \\

Energy conservation was applied to reconstruct the kinematics of the top quarks. The longitudinal component of the neutrino momentum ($p^\nu_z$) was then determined, together with the choice for the jet and lepton combination, by minimizing the $\chi ^{2}$ as defined in Equation~\ref{chi2}:\\
\begin{equation}
\label{chi2} 
\resizebox{0.9\columnwidth}{!}{$
\chi ^{2} = \frac{ \left( m^{\textrm{\tiny reco}}_{j_a l_a l_b} - m_t \right) ^2} {\sigma ^2 _t} + \frac{\left( m^{\textrm{\tiny reco}}_{j_b l_c \nu} -m_t \right) ^2} {\sigma ^2 _t}
\frac{\left( m^{\textrm{\tiny reco}}_{l_c \nu} -m_W \right) ^2} {\sigma ^2 _W} + \frac{\left( m^{\textrm{\tiny reco}}_{ l_a l_b} -m_Z \right) ^2} {\sigma ^2 _Z}
$}
\end{equation}
where $m^{\textrm{\tiny reco}}_{j_a l_a l_b}$, $m^{\textrm{\tiny reco}}_{j_b l_c \nu}$, $m^{\textrm{\tiny reco}}_{l_c \nu}$, and $m^{\textrm{\tiny reco}}_{ l_a l_b}$ are the reconstructed masses of the top quark decaying through the FCNC decay, the top quark with SM decay, the $W$-boson from the top quark with SM decay, and of the $Z$-boson from the top quark with FCNC decay respectively. The following values were used for the constraints: $m_t = 172.5$~GeV, $m_W = 80.4$~GeV, $m_Z = 91.2$~GeV, $\sigma _t = 14$~GeV, $\sigma _W = 10$~GeV and $\sigma _Z = 3$~GeV. For each jet and lepton combination (the Z-candidate built from two leptons of same flavor and opposite charges) the $\chi ^2$ minimization gives the most probable value for $p^\nu _z$. From all combinations, the one with the minimum $\chi ^2$ was chosen, along with the corresponding $p^\nu _z$ value. The reconstructed $W$-boson mass was then required to be within 30 GeV of the $W$-mass peak and the reconstructed top quark masses to be within 40 GeV of the top quark mass. Table~\ref{tab:yields} shows the number of data events, the number of expected background (described in Section~\ref{bkg}) events and the signal efficiency after the final event selection.\\

\begin{table}
\centering
\caption{Number of selected data events, expected number of background events ($ZZ$ and $WZ$ events estimated from MC simulation samples and DD estimations for 1 and 2+3 fake leptons) and the estimated signal efficiency (multiplied by the corresponding $W$- and $Z$-bosons' BRs), after the final selection. The corresponding statistical uncertainties are also shown.}
\label{tab:yields}       
\resizebox{0.75\columnwidth}{!}{%
\begin{tabular}{ll}
\hline\noalign{\smallskip}
$ZZ$ and $WW$ & $2.4 \pm 0.3$   \\
1+2+3 fake leptons (DD) & $0.0 ^{+1.8} _{-0.0}$ \\
\noalign{\smallskip}\hline\noalign{\smallskip}
Expected background & $2.4 ^{+1.8} _{-0.3}$ \\
\noalign{\smallskip}\hline\noalign{\smallskip}
data & 2 \\
\noalign{\smallskip}\hline\noalign{\smallskip}
Signal efficiency & $\left( 0.209 \pm 0.004 \right)$ \% \\
\noalign{\smallskip}\hline
\end{tabular}
}
\end{table}
%

\subsection{Background estimation}
\label{bkg}
The main background comes from $ZZ$ and $WZ$ events which have three charged leptons in the observed final state, and were estimated with MC simulation. Backgrounds in which a jet is reconstructed as a lepton (such as $WW$ and leptonic $t\bar{t}$ events) were estimated at the preselection level with simulation samples and at the final selection with DD methods. Events with two (such as $W$+jets and single lepton $t\bar{t}$ events) or three (such as QCD multi-jet and hadronic $t\bar{t}$ events) jets faking leptons, were estimated by DD methods. \\

For the estimation of the $Z$+jets in the FCNC signal region, at the preselection level a DD method was developed, similar to the one used for the ATLAS $t\bar{t}$ cross-section measurement \cite{bib:tT}. This method uses a control region in the ($E_T^{miss}, m_{ll}$) plane, by selecting events with two leptons, $E_T^{miss} \leq 20$~GeV and $|m_Z-m_{ll}^{\textrm{\tiny reco}}|<15$~GeV. The $Z$+jets estimate in the FCNC signal region is then simply the number of simulation $Z$+jets events in the signal region scaled by the ratio of data events (reduced by the MC simulation expectation of other backgrounds) to the number of simulation $Z$+jets events, both counted in the control region, as shown in Equation~\ref{eq:z}.\\
\begin{equation}
\label{eq:z} 
\resizebox{0.9\columnwidth}{!}{$
\left[ N_{\textrm{\tiny Z+jets}}^{\textrm{\tiny Data}} \right] _{\textrm{\tiny Signal Region}} =
\left[   \frac{N^{\textrm{\tiny Data}} - N_{\textrm{\tiny other bkgs}}^{\textrm{\tiny MC}}  }{N_{\textrm{\tiny Z+jets}}^{\textrm{\tiny MC}} }  \right]_{\textrm{\tiny Control Region}}
\times \left[ N_{\textrm{\tiny Z+jets}}^{\textrm{\tiny MC}}  \right]_{\textrm{\tiny Signal Region}}
$}
\end{equation}
The remaining backgrounds, at the preselection level, with one fake lepton were estimated using MC simulation. \\

To estimate the contribution from the QCD multi-jet, $W$+jets, single top and $t\bar t$ single lepton decay events, in which two or three jets were reconstructed as leptons (2+3 fake leptons) a DD method was developed. Due to the requirement that two leptons should have the same flavor and opposite charges, the yield from these backgrounds can be extrapolated from the number of observed data events with three leptons with the same charge. Taking into account the possible charge and flavor combinations, there are 36 combinations of three leptons, from which two must have the same flavor and opposite charges and 16 combinations of three leptons with the same charge. The extrapolation factor is thus f = 36/16 . No data event passed the selection after requiring three leptons with the same charge. Using MC simulation samples, $0.0^{+1.6} _ {-0.0}$ events were predicted, which is consistent with the DD method.\\

For events with one jet misidentified as a lepton and two real leptons with opposite charges (one fake
lepton) different methods were developed. Not applying the $E_T^{miss}$ requirement would give the same number of data events. 
This result is a sign of a very low contamination from $Z$+jets events, which concentrate in the low $E_T^{miss}$ region. 
For the $t\bar t$ dileptonic, $WW$, and single top $Wt$ production channels, background estimates were made by
comparing the event selection with all three leptons passing the tight criteria, as in the reference analysis,
with the event selection with two tight leptons and one lepton with loosened cuts. 
No additional events were not observed, indicating that the background from these channels has a very low yield.
 The predicted number of events from these backgrounds using
simulation samples are $0.0^{+2.6}_{-0.0}$, $0.0^{+0.02}_{-0.00}$, and $0.0^{+0.02}_{-0.00}$ for $Z$+jets, $t\bar t$ dilepton and single top $Wt$-channel production, respectively. These results further confirm the DD method.

\subsection{Limit evaluation}
\label{limit}
Good agreement between data and background yields was observed, as shown in Table~\ref{tab:yields} for the final
selection level. No evidence for the $t \rightarrow qZ$ decay was found, and 95\% CL upper limits on the number
of signal events were derived using the modified frequentist likelihood method \cite{bib:freq}. These limits
were evaluated using 10$^5$ pseudo-experiments of the expected signal and background samples and were
converted into upper limits on the corresponding branching ratio (BR) using the approximate NNLO calculation, and uncertainty, for the $t\bar t$ cross section ($\sigma_{t\bar t} = 165^{+11}_{-16}$~pb), and constraining $BR(t\rightarrow bW) = 1 - BR(t\rightarrow qZ)$. Statistical and systematic uncertainties were taken into account. Systematic uncertainty effects were implemented assuming a Gaussian distribution. Table~\ref{tab:limit} shows the observed and expected limits together with the $\pm 1 \sigma$ bands. 
\begin{table}
\centering
\caption{The observed 95\% CL upper limits on the FCNC top quark decay $t \rightarrow qZ$ BR are shown. The expected values with the $\pm1 \sigma$ limit, which include both the statistical and the systematic uncertainty contributions, are also presented.}
\label{tab:limit}     
\resizebox{0.95\columnwidth}{!}{%
\begin{tabular}{ccccc}
\hline\noalign{\smallskip}
  & observed & $\left( -1 \sigma \right)$ & expected &   $\left( +1 \sigma \right)$ \\
\noalign{\smallskip}\hline\noalign{\smallskip}
stat. uncertainty only & $1.06 $\% & $0.78$ \% & $1.22$ \% & $1.94$ \% \\
stat. + syst. uncertainties & $1.13 $\% & $0.83$ \% & $1.30$ \% & $2.09$ \% \\
\noalign{\smallskip}\hline
\end{tabular}
}
\end{table}
%

\section{Conclusions}
\label{conclusions}
A search for the $t \rightarrow qZ$ decay process was performed by studying top quark pair production with one top quark decaying according to the SM and the other according to the FCNC: $t\bar{t} \rightarrow bWqZ$. No evidence for such a signal was found. An observed limit at 95\% CL on the $t \rightarrow qZ$  FCNC top quark decay BR was set at BR$\left( t \rightarrow qZ \right) < 1.1 $ \%, assuming BR$\left( t \rightarrow bW \right)$+BR$\left( t \rightarrow qZ \right)$ $= 1$. The observed limit is compatible with the expected sensitivity, assuming that the SM describes correctly the data, BR$\left( t \rightarrow qZ \right)$ $<1.3$\%.
\begin{table}
\centering
\caption{Present experimental 95\% CL upper limits on the BR of the FCNC top quark decay channels (list is not exhaustive).}
\label{tab:compare}       
\resizebox{0.9\columnwidth}{!}{%
\begin{tabular}{ccccc}
\hline\noalign{\smallskip}
  & LEP & HERA & Tevatron & ATLAS  \\ 
  BR$\left( t \rightarrow qZ \right)$ & $7.8$ \% & $49$ \% $\left( tuZ \right)$ & $3.2$ \% & $1.1$ \%  \\ 
\noalign{\smallskip}\hline
\end{tabular}
}
\end{table}
Other experiments at LEP, HERA, and Tevatron accelerators have set experimental limits on the BR of this FCNC top quark decay. See Table~\ref{tab:compare}. The limit presented here is the best direct measurement available.

\end{document}